\documentclass[preprint, authoryear]{elsarticle}

\usepackage{setspace}
\usepackage{mleftright}
\usepackage{mathtools}
\usepackage[normalem]{ulem}
\usepackage{comment}
\usepackage{amsmath, amssymb, amsfonts}
\usepackage{bm}
\usepackage{epstopdf}
\usepackage{array}
\usepackage{dcolumn}
\usepackage{longtable}
\usepackage{natbib}
\usepackage{color}
\usepackage{subfigure}
\usepackage{url}
\usepackage{graphicx}
\usepackage{stfloats}
\usepackage[labelfont=bf,labelsep=period]{caption}

\usepackage{ascmac}
\usepackage[colorlinks=true, linkcolor=blue, citecolor=blue, urlcolor=blue]{hyperref}

\renewcommand{\figurename}{Fig.}

\journal{Journal of Theoretical Biology}

\begin{document}

\begin{frontmatter}

\title{
       An energy landscape-based theoretical framework for understanding the emergence of functions in a living system under the dynamical component interaction
      }

\author[1,2]{Ryunosuke Suzuki}
\ead{suzuki.ryunosuke.55s@st.kyoto-u.ac.jp}

\author[1,2,3]{Taiji Adachi\corref{cor1}}
\ead{adachi@infront.kyoto-u.ac.jp}

\cortext[cor1]{Corresponding author}

\affiliation[1]{
                organization={Laboratory of Biomechanics, Department of Biosystems Science, Institute for Life and Medical Sciences, Kyoto University},
                addressline={53 Shogoin-Kawahara-cho}, 
                city={Sakyo-ku},
                state={Kyoto},
                postcode={606-8507}, 
                country={Japan}}

\affiliation[2]{
                organization={Department of Micro Engineering, Graduate School of Engineering, Kyoto University},
                addressline={53 Shogoin-Kawahara-cho}, 
                city={Sakyo-ku},
                state={Kyoto},
                postcode={606-8507}, 
                country={Japan}}

\affiliation[3]{
                organization={Department of Mammalian Regulatory Network, Graduate School of Biostudies, Kyoto University},
                addressline={53 Shogoin-Kawahara-cho}, 
                city={Sakyo-ku},
                state={Kyoto},
                postcode={606-8507}, 
                country={Japan}}

\begin{abstract}
In a living system composed of interacting components such as molecules, cells, and tissues, each component often adaptively changes its internal states in response to interactions with its surrounding components. For example, individual tissues exhibit component-level adaptive behavior, such as growth and remodeling, in response to their mechanical interactions, resulting in the emergence of functions of a multi-tissue system. Along with the adaptive behavior of the components, their interactions exhibit dynamical changes, which strongly influence the emergence of system functions. To understand how the emergence of system functions occurs from such dynamical interactions due to component-level adaptive behavior, this study proposes a theoretical framework that formulates the dynamics of interactions among components due to the adaptive behavior of individual components. For modeling the adaptive internal state changes, we assign an energy landscape and its associated energy rate landscape for each component, leading to the generalized gradient flow model of adaptive behavior. Then, we represent interaction dynamics based on temporal changes in these energy and energy rate landscapes by formulating temporal changes in the environmental states of each component due to the adaptive behavior of individual components. Through case studies using simplified models of mechanically interacting tissues under morphological changes, our theoretical framework demonstrates that temporal changes in applied forces due to morphological changes of individual tissues determine the emergence of system functions. These findings highlight that expressing interaction dynamics based on temporal changes in energy and energy rate landscapes offers a powerful theoretical framework for understanding how system functions emerge from component-level adaptive behavior.
\end{abstract}


\begin{keyword}
   Living system \sep 
   Energy landscape \sep 
   Generalized gradient flow \sep 
   Adaptive behavior \sep 
   Interaction
\end{keyword}

\end{frontmatter}

\section{Introduction}\label{sec1}
Living systems, which consist of components such as molecules, cells, and tissues, exhibit functions that emerge from interactions among components as well as the capabilities of individual components. At the molecular scale, interactions among biomolecules, such as dynamical complex formation \citep{beusch_understanding_2024, borsley_membrane_2024} and allosteric binding \citep{wodak_allostery_2019, hofmann_all_2023, wu_allosteric_2024}, govern the structural configuration and activity of molecular assemblies. At the cellular scale, interactions among cells mediated by forces and biochemical signals regulate tissue dynamics such as morphogenesis \citep{vignes_mechanical_2022, su_cellcell_2024} and drive functional mechanisms such as suppression of tumor expansion \citep{van_luyk_cell_2024, kim_entosis_2024}.

In response to these interactions in some living systems, individual components adaptively change their internal states, leading to the emergence of functions in an integrated living system beyond the capabilities of individual living components. For example, when living tissues interact mechanically, individual tissues grow and remodel themselves in response, resulting in the emergence of the functional multi-tissue system \citep{felsenthal_mechanical_2017, ambrosi_growth_2019, killian_growth_2022}. Moreover, the dynamics of interactions among living components affect the functions of an integrated living system through component-level adaptive behavior. For example, in the phenomenon known as the ball-and-socket ankle, which is observed in ankle joints during developmental processes, changes in the interactions among bones result in a shift from a typical hinge joint morphology to a spherical morphology \citep{stevens_ball-and-socket_2006}, leading to the alternation of the functional range of motion of the joints \citep{jastifer_ball_2017}. Thus, the dynamical interactions among living components play a crucial role in controlling the emergence of functions of an integrated living system through the adaptive behavior of individual components.

A successful approach for characterizing the adaptive behavior of living components is to build a theory based on energy landscapes. These energy landscapes symbolize the Waddington landscape, which reflects the surrounding environment of a living component \citep{waddington_strategy_1957, waddington_nature_1961}. In this analogy, a position on the energy landscape corresponds to the adaptively changing internal states of a living component, whereas the adaptive behavior and temporal environment changes correspond to the rolling of a ball and temporal landscape changes, respectively.

Previous studies have formulated adaptive behavior of living components using energy landscapes \citep{furusawa_dynamical-systems_2012, matsushita_homeorhesis_2020}. Furthermore, by assigning energy landscapes for an integrated living system in addition to its individual components, a theory has incorporated system-level adaptive behavior, which reveals a correspondence between a system-level energy landscape and landscapes of components \citep{horiguchi_cellular_2023}. Consequently, this theory has described the dynamics of energy landscapes of components associated with system-level adaptive behavior.

Considering interactions among living components in energy landscape-based approaches is expected to capture the emergence of functions of an integrated living system controlled by these interactions. When considering interactions among living components, changes in the surrounding environment of each component correspond to interaction dynamics due to the adaptive behavior of individual components. Therefore, to incorporate interactions among components into these energy landscape-based approaches, it is necessary to formulate the dynamics of energy landscapes as a result of the adaptive behavior of individual components, especially without considering system-level adaptive behavior. By developing such a framework, the effects of interaction dynamics due to the adaptive behavior of individual components on the emergence of system functions can be analyzed.

This study aims to propose a theoretical framework that embodies the dynamical interactions among living components based on their energy landscapes. By formulating adaptive internal state changes of individual living components and interaction dynamics due to the adaptive behavior, this framework embodies energy landscape changes for each living component. Additionally, through case studies using simplified models of living systems comprising mechanically interacting tissues, we demonstrate that the proposed framework enables us to analyze the effects of interaction dynamics on the emergence of system functions, especially without explicitly defining system-level adaptive behavior.

\section{Theoretical framework}
\begin{figure}[h]
  \centering
  \includegraphics[width=\columnwidth]{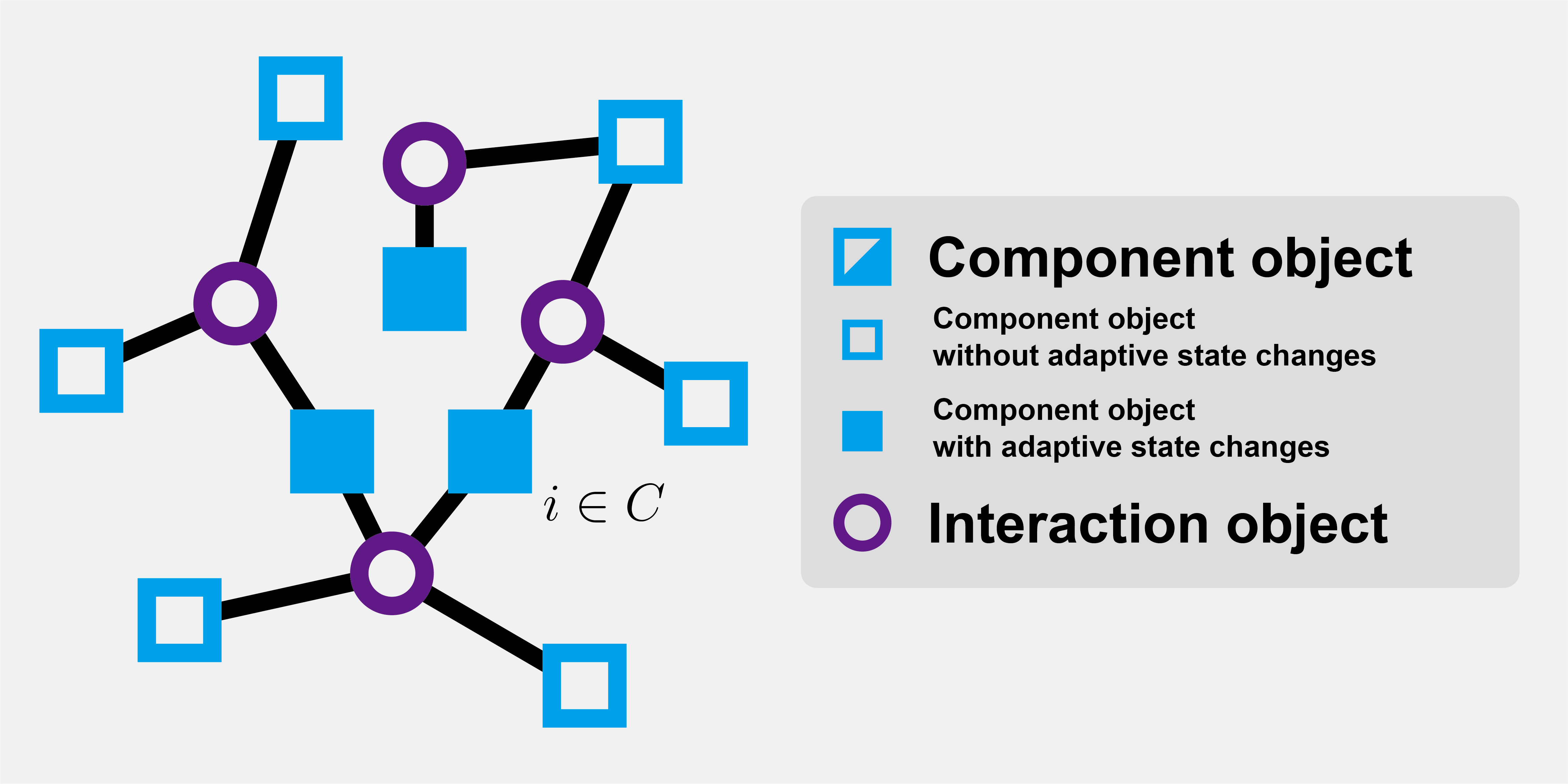}
  \caption{Graph composed of component and interaction objects.}
  \label{fig: system}
\end{figure}

In this chapter, to embody the dynamical interactions among living components within a living system, we formulate the adaptive behavior of individual living components and the dynamics of energy landscapes due to the adaptive behavior. First, we define graphs composed of component and interaction objects as shown in \figurename~\ref{fig: system}. Here, component objects are introduced to describe the adaptive behavior of individual living components (\figurename~\ref{fig: system}, squares), and interaction objects are introduced to describe the dynamics of interactions among components (\figurename~\ref{fig: system}, circles). Next, we adopt the framework of the generalized gradient flow to the formulation of adaptive behavior of individual living components, given that it is well-suited for formulating temporal changes in the internal states of living components viewed as dissipative behavior \citep{horiguchi_cellular_2023, mielke_gradient_2011, peletier_variational_2014}. This framework assigns an energy function for each component object to define its energy landscape. The adaptive behavior of each component object is described by the rolling of a ball in its energy landscape, where a position corresponds to its internal states. Moreover, to express the temporal changes in energy functions of individual component objects due to their adaptive behavior, we formulate temporal environment changes surrounding each component object as functions of adaptive internal state changes of component objects connected through the same interaction object. Such environment changes determine the actual internal state changes of the corresponding component object. As a result, dynamical interactions among component objects, due to their adaptive behavior, are described based on temporal changes in their energy landscapes.

\subsection{Formulation of adaptive behavior of individual component objects}
To formulate the adaptive behavior of individual component objects, for a solid square component object $i \in C$ in \figurename~\ref{fig: system}, where $C$ is a set of component objects, $x_i(t) \in S^{x}_{i}$ denotes its internal states that adaptively change over time. To distinguish adaptively changing internal states from the other internal states, the non-adaptively changing internal states and total internal states are denoted as $y_i(t) \in S^{y}_{i}$ and $s_i(t)(=(x_i(t),y_i(t))) \in S_{i}(=S^{x}_{i} \times S^{y}_{i})$, respectively. Here, $S^{x}_{i}$, $S^{y}_{i}$, and $S_{i}$ are manifolds that form subspaces of Euclidean space. In contrast, open square component objects in \figurename~\ref{fig: system} are assumed to behave non-adaptively, which is described as $s_i(t)=y_i(t)$. A rate of the adaptive internal states of each component object is denoted as $\underline{\dot x}_{i}(t) \in T_xS^{x}_{i}$, which serves as a basis for determining actual rates of internal states through dynamics of interactions among component objects. These rates are denoted as $\dot x_i(t) \in T_xS^{x}_{i}$, $\dot y_i(t) \in T_{y}S^{y}_i$, and $\dot s_i(t) \in T_sS_i$, where $T_xS^{x}_{i}$, $T_{y}S^{y}_i$, and $T_sS_i\ \mleft(= T_xS^x_i \times T_{y}S^{y}_i\mright)$ are the tangent spaces of $S^{x}_{i}$, $S^{y}_{i}$, and $S_{i}$, respectively. Note that $\underline{\dot x}_{i}(t)=0$ is assumed for open square component objects because their internal states change non-adaptively over time.

\begin{figure}[h]
  \centering
  \includegraphics[width=\columnwidth]{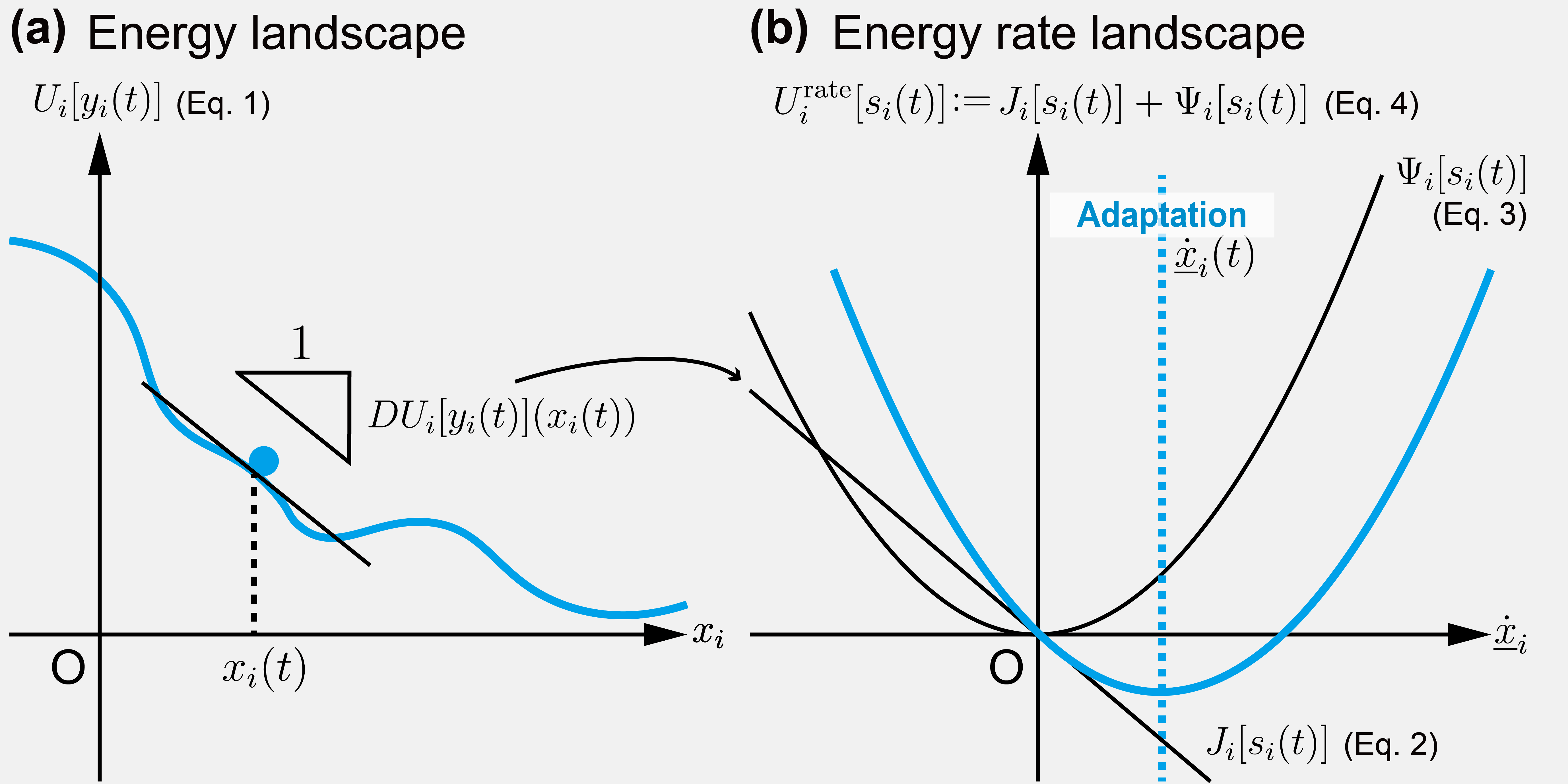}
  \caption{Energy and energy rate landscapes of a component object $i \in C$.
  (a) Energy landscape formed by the energy function $U_i[y_i(t)]$.
  (b) Energy rate landscape formed by the function $U^\text{rate}_i[s_i(t)]$.
  }
  \label{fig: gradient}
\end{figure}

For solid square component objects, which adaptively change their internal states, the rate $\underline{\dot x}_{i}(t) \in T_xS^{x}_{i}$ is formulated as dissipative behavior based on the framework of the generalized gradient flow \citep{horiguchi_cellular_2023, peletier_variational_2014}. First, for a solid square component object $i \in C$, its energy function with the domain $S^x_i$ is defined as
\begin{align}
  U_i[y_i(t)]:
  S^{x}_{i}  \longrightarrow  \mathbb{R};\ 
  x_i  \longmapsto  U_i[y_i(t)](x_i), 
  \label{eq: energy function}
\end{align}
where $``[y_i(t)]"$ means that the energy function $U_i[y_i(t)]$ depends on the non-adaptively changing internal state $y_i(t)$, forming an energy landscape of the component object $i$ (\figurename~\ref{fig: gradient}(a), blue curve).

Next, to formulate a rate of the adaptive internal states of a component object as the generalized gradient flow of an energy landscape, energy change rate and energy dissipation rate functions of $\underline{\dot x}_{i}$ are defined as
\begin{align}
  J_i[s_i(t)]&:
  T_xS^{x}_{i}  \longrightarrow  \mathbb{R};\ 
  \underline{\dot x}_{i}  \longmapsto  \langle 
                                        DU_i[y_i(t)](x_i(t)),\underline{\dot x}_{i} 
                                       \rangle,
  \label{eq: energy gradient function}
  \\
  \Psi_i[s_i(t)]&:
  T_xS^{x}_{i}  \longrightarrow  \mathbb{R};\ 
  \underline{\dot x}_{i}  \longmapsto  \dfrac{1}{2} \langle 
                                                      \underline{\dot x}_{i}, \omega_i(s_i(t))\underline{\dot x}_{i} 
                                                    \rangle, 
  \label{eq: cost function}
\end{align}
which are illustrated by a black line and parabola in \figurename~\ref{fig: gradient}(b). Here, when $T^*_xS^{x}_{i}$ denotes the cotangent space of the set of adaptively changing internal states $S^{x}_{i}$, $\langle \tau,\xi \rangle=\langle \xi,\tau \rangle \in \mathbb{R}$ is a contraction operation that returns a scalar quantity (an energy change rate in Eq.~\eqref{eq: energy gradient function} or an energy dissipation rate in Eq.~\eqref{eq: cost function}) for $\tau \in T_xS^{x}_{i}$ and $\xi \in T^*_xS^{x}_{i}$. In addition, $DU_i[y_i(t)]: S^{x}_{i} \longrightarrow T^*_xS^{x}_{i}$ is a function that returns a derivative $DU_i[y_i(t)](x_i)$ of $U_i[y_i(t)]$ at a given input $x_i$ (\figurename~\ref{fig: gradient}(a), gradient of the black tangent line). Furthermore, the function $\omega_i: S_{i} \longrightarrow T^*_xS^{x}_{i} \times T^*_xS^{x}_{i}$ is a function of the total internal state $s_{i}(t)$ that provides the weight function $\omega_i(s_i(t)): T_xS^{x}_{i} \longrightarrow T^*_xS^{x}_{i}$ that describes the energy dissipation rate for the adaptive behavior, ensuring that the energy dissipation rate function $\Psi_i[s_i(t)]$ is positive definite.

The sum of the energy change rate function $J_i[s_i(t)]$ and the energy dissipation rate function $\Psi_i[s_i(t)]$, that is, the net energy change rate function accounting for energy dissipation is defined as
\begin{align}
  U^\text{rate}_i[s_i(t)] := J_i[s_i(t)] + \Psi_i[s_i(t)]
  \label{eq: Urate}
\end{align}
which is illustrated by a blue parabola in \figurename~\ref{fig: gradient}(b), defining the condition for the temporal rate $\underline{\dot x}_{i}(t)$ of the adaptive internal state changes to be the generalized gradient flow as a minimization condition
\begin{align}
  \underline{\dot x}_{i}(t)\ \text{minimizes}\ U^\text{rate}_i[s_i(t)].
  \label{eq: gradient flow}
\end{align}
Thus, the function $U^\text{rate}_i[s_i(t)]$ forms an energy rate landscape for the component object $i$ (\figurename~\ref{fig: gradient}(b), blue parabola), whose bottom determines temporal change rate $\underline{\dot x}_{i}(t)$ of the adaptive internal state changes (\figurename~\ref{fig: gradient}(b), blue dotted line). The interaction dynamics converts this rate into the actual rate ${\dot x}_i(t)$, which in turn allows the internal state $x_i(t)$ to evolve within the energy landscape.

\subsection{Formulation of dynamical interactions among component objects}
\begin{figure}[h]
  \centering
  \includegraphics[width=\columnwidth]{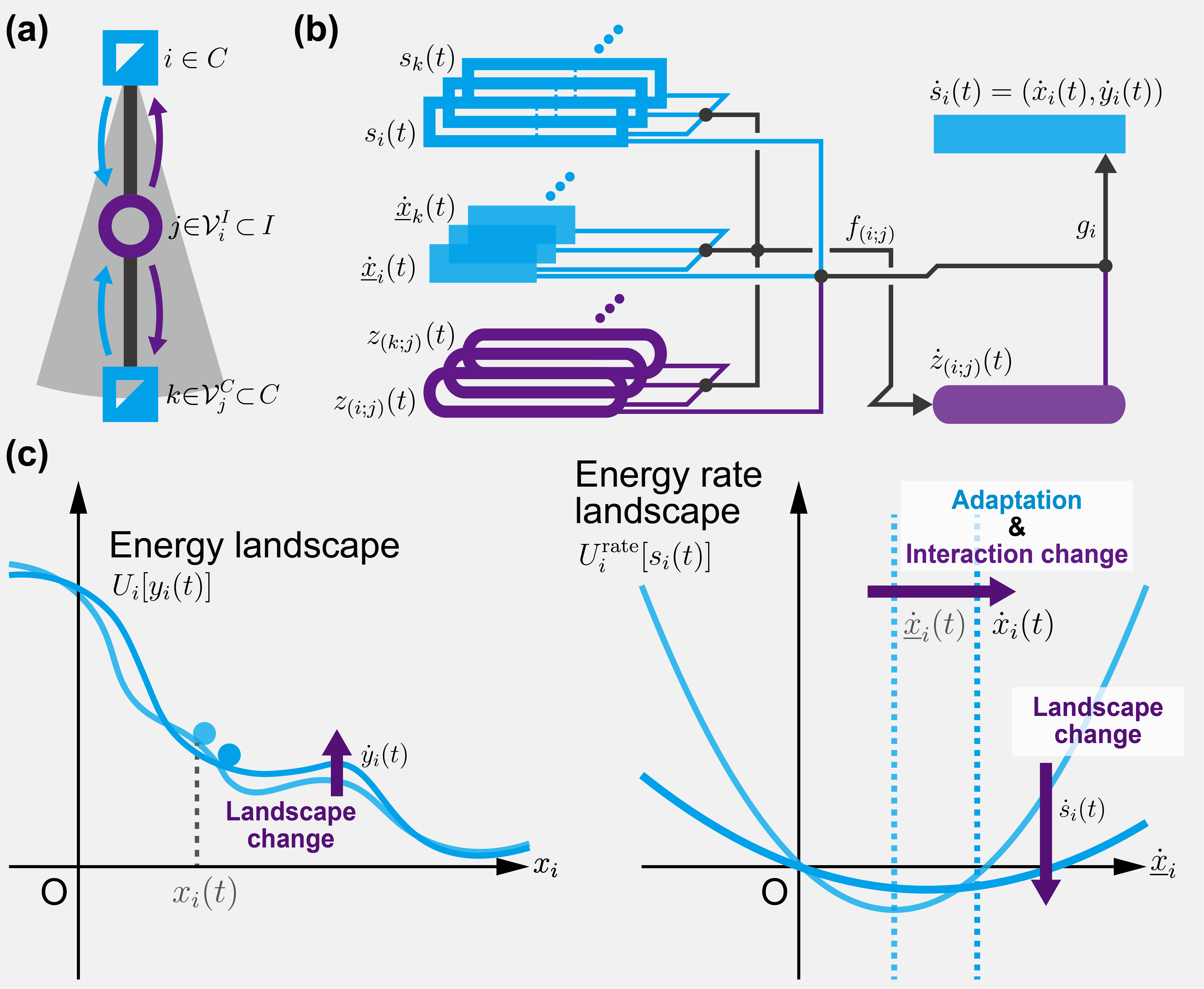}
  \caption{
    Formulation of dynamical interactions among component objects. 
  (a) Adjacency of a component object $i \in C$ and an interaction object $j \in I$. 
  (b) Functions $f_{(i;j)}$ to formulate environmental state changes and $g_i$ to formulate internal state changes. 
  (c) Landscape changes caused by interaction dynamics due to adaptive internal state changes.
  }
  \label{fig: view}
\end{figure}

To formulate dynamical interactions among component objects due to their adaptive behavior, let $I$ denote a set of interaction objects, $\mathcal{V}^{I}_{i} \subset I$ a set of interaction objects adjacent to a component object $i \in C$, and $\mathcal{V}^{C}_{j} \subset C$ a set of component objects adjacent to an interaction object $j \in I$. With these objects, a situation in which component objects $i$ and $k$ are connected through the same interaction object $j$ is expressed as $j \in \mathcal{V}^{I}_{i},\ k \in \mathcal{V}^{C}_{j}$ (\figurename~\ref{fig: view}(a)). For an interaction object $j \in I$ and a component object $k \in \mathcal{V}^{C}_{j}$ (including the component object $i$), $S^z_{(k;j)}$ denotes a manifold that forms a subspace of Euclidean space, and its component $z_{(k;j)}(t) \in S^z_{(k;j)}$ denotes environmental states of the component object $k$ associated with the interaction object $j$ at time $t$.

To formulate temporal environment changes of a component object $i$ due to adaptive behavior of individual component objects, we introduce a function $f_{(i;j)}\ (j\in\mathcal{V}^{I}_{i})$, which maps the total internal state $s_k(t)$, a rate of adaptive internal state $\underline{\dot x}_{k}(t)$, and the environmental state $z_{(k;j)}(t)$, where $k \in \mathcal{V}^{C}_{j}$, to a temporal rate of environmental state $\dot z_{(i;j)}(t)$ of a component object $i$, denoted as
\begin{align}
  f_{(i;j)}:
 \underset{k\in\mathcal{V}^{C}_{j}}
  \prod (S_{k} 
        \times T_xS^{x}_{k} 
        \times S^z_{(k;j)})  &\longrightarrow  T_{z}S^z_{(i;j)};
  \notag\\
  (s_k(t), 
   \underline{\dot x}_{k}(t), 
   z_{(k;j)}(t))_{k\in\mathcal{V}^{C}_{j}}  &\longmapsto  \dot z_{(i;j)}(t)
  \label{eq: function f}
\end{align}
(\figurename~\ref{fig: view}(b)). Furthermore, to formulate actual internal state changes of a component object $i \in C$ due to changes in its environmental state, we introduce a function $g_{i}$, which maps the total internal state $s_i(t)$, the rate of adaptive internal state $\underline{\dot x}_{i}(t)$, and the environmental state $z_{(i;j)}(t)$ as well as its temporal rate $\dot z_{(i;j)}(t)$, where $j\in\mathcal{V}^{I}_{i}$, to a temporal rate of the internal state $\dot s_i(t)$, denoted as
\begin{align}
  g_{i}:
 S_i 
  \times T_xS^{x}_{i} 
  \times \underset{j\in\mathcal{V}^{I}_{i}} \prod 
         \mleft(S^z_{(i;j)} \times T_{z}S^z_{(i;j)}\mright)
    &\longrightarrow  T_sS_i\mleft(= T_xS^x_i \times T_{y}S^{y}_i\mright); 
  \notag\\
 \mleft(s_i(t),
         \underline{\dot x}_{i}(t),
         z_{(i;j)}(t), 
         \dot z_{(i;j)}(t)\mright)_{j\in\mathcal{V}^{I}_{i}} 
    &\longmapsto  \dot s_{i}(t)\mleft(=(\dot x_{i}(t), \dot y_{i}(t))\mright)
  \label{eq: function g}
\end{align}
(\figurename~\ref{fig: view}(b)). Thus, the set of functions leads to the temporal changes in the internal state $x_{i}(t)$, the energy landscape formed by $U_i[y_i(t)]$, and the energy rate landscape formed by $U^\text{rate}_i[s_i(t)]$ (\figurename~\ref{fig: view}(c), purple arrow).

Consequently, in addition to the adaptive behavior of individual component objects, environmental and internal state changes ($\dot z_{(i;j)}(t)$ and $\dot s_i(t)$), each corresponding to adaptive internal state changes ($\underline{\dot x}_{k}(t)$), were formulated. Thus, this theoretical framework embodies the dynamics of interactions among living components due to the adaptive behavior of individual living components as energy landscape changes in response to a rate $\dot y_i(t)$ and energy rate landscape changes in response to a rate $\dot s_i(t)$.

\newcommand{\IntOne}{GA}
\newcommand{\IntTwo}{EM}
\newcommand{\IntThree}{ErM}

\section{Case studies}
In this chapter, we show that our formulation of the dynamical interactions among living components, based on temporal changes in the energy and energy rate landscapes respectively formed by the functions $U_i[y_i(t)]$ and $U^\text{rate}_i[s_i(t)]$, allows us to analyze the influence of the dynamics of interactions among components on the emergence of functions of a living system without explicitly defining the adaptive behavior of an integrated living system. To achieve this, we define three different conditions for the functions $f_{(i;j)}$ and $g_i$ ($i \in C, j \in \mathcal{V}^{I}_{i}$), which represent interactions among component objects, and examine how dynamics of component interactions affects the emergence of system functions under each condition.

The first interaction condition (\textbf{\IntOne{}}: Global adaptive interaction) explicitly defines the adaptive behavior of an integrated living system, which is associated with the functions $f_{(i;j)}$ and $g_i$. In contrast, instead of the adaptive behavior of the integrated living system, the second (\textbf{\IntTwo{}}: Energy landscape modified interaction) and third (\textbf{\IntThree{}}: Energy rate landscape modified interaction) interaction conditions define the functions $f_{(i;j)}$ and $g_i$ such that the dynamics of the energy landscape and the energy rate landscape exhibits the emergence of functions of the system, respectively. By comparing the dynamics of the energy landscape between \textbf{\IntOne{}} and \textbf{\IntTwo{}}, as well as the dynamics of the energy rate landscape between \textbf{\IntOne{}} and \textbf{\IntThree{}}, we demonstrate that our theory can analyze the influence of the interaction dynamics on the emergence of functions of the integrated living system even when the interaction conditions do not explicitly define adaptive behavior of the system.

\subsection{Set up of adaptive state changes of living components}
In this section, we define the internal and environmental states of component objects that constitute a system and the adaptive internal state changes of individual component objects in response to their interactions. As an example of adaptive behavior, these case studies consider a scenario where individual living tissues undergo morphological changes in response to their mechanical interactions. For simplicity, we limit the number of component and interaction objects to two and one, denoted as $C = \{1, 2\}$ and $I = \{0\}$, respectively.

We model the component objects 1 and 2 constituting the system as rod-shaped living components that adaptively change their cross-sectional areas $a_i(t)$ ($i = 1,2$) in response to forces (\figurename~\ref{case study}(a)). Each component object $i$ ($i = 1,2$) is a linearly elastic rod with a constant natural length $l_i$ and Young modulus $e_i$. These rods are connected in series at their respective natural lengths, with both ends fixed to rigid walls separated by a distance $l := l_1 + l_2$.

Each component object has its cross-sectional area $a_i(t)$ ($i=1,2$) as the adaptively changing internal state $x_i(t)$, denoted as
\begin{align}
  x_i(t) = a_i(t)  \in  S^x_i=\mathbb{R}_{+},
  \label{eq: cs x_i}
\end{align}
where $\mathbb{R}_{+} (\subset \mathbb{R})$ represents the set of positive real numbers. Additionally, each component object has the internal state $y_i(t)$ that does not change over time due to adaptive behavior. This state consists of the axial strain $\epsilon_i(t)$, its gradient with respect to the cross-sectional area ${\partial \epsilon_i}/{\partial a_i}(t)$, and the weight $\omega_i(t)(:=\omega_i(s_i(t)))$ of the energy dissipation rate in adaptive behavior, denoted as
\begin{align}
  y_i(t)=
  \mleft(\epsilon_i(t), 
         \dfrac{\partial \epsilon_i}{\partial a_i}(t), 
         \omega_i(t) 
  \mright)
  \in S^y_i=\mathbb{R}^2 \times \mathbb{R}_{+}.
  \label{eq: cs alpha_i}
\end{align}
Consequently, the total internal state $s_i(t)$ of the component object $i$ at time $t$ is given by
\begin{align}
  s_i(t)
  = (x_i(t),y_i(t))
  = \mleft(a_i(t), 
           \epsilon_i(t), 
           \dfrac{\partial \epsilon_i}{\partial a_i}(t), 
           \omega_i(t)
    \mright)
    \in S_i
        = \mathbb{R}_{+} 
          \times \mathbb{R}^2 
          \times \mathbb{R}_{+}.
  \label{eq: cs s_i}
\end{align}

The connection between the component objects is under an external force $p$ in their axial direction, as shown in \figurename~\ref{case study}(a). As a result of the mechanical interaction mediated by interaction object 0, each component object experiences a reaction force $r_i(t)$ ($i=1,2$) from the rigid wall. Here, under the positive direction of the external and reaction forces $p$ and $r_i(t)$ respectively defined as pointing from the component object 2 toward the component object 1 and from the wall toward the component object $i$, the equilibrium condition among these forces is given by $r_1(t) - r_2(t) + p = 0$. Additionally, the compatibility condition for the elastic deformation in the axial direction at the connection, expressed as ${r_1(t)}/{(e_1 a_1(t)/l_1)} + {r_2(t)}/{(e_2 a_2(t)/l_2)} = 0$. Consequently, the reaction force $r_i(t)$ and its gradient with respect to the cross-sectional area ${\partial r_i}/{\partial a_i}(t)$ are given by  
\begin{align}
  r_i(t)
&:= (-1)^i 
    \dfrac{{e_ia_i(t)}/{l_i}}
          {\sum_{k\in C}
           \mleft({e_ka_k(t)}/{l_k}\mright)}\ p,
  \label{eq: cs real Reaction force}\\ 
  \dfrac{\partial r_i}{\partial a_i}(t)
&:= (-1)^i 
    \dfrac{\mleft({e_i}/{l_i}\mright) 
           \mleft({e_j}a_j(t)/{l_j}\mright)}
          {\mleft(\sum_{k\in C}
                  \mleft({e_ka_k(t)}/{l_k}\mright)
           \mright)^2}\ p
  \quad (j \neq i).
  \label{eq: cs real Reaction force gradient}
\end{align}

To define temporal changes in strain $\epsilon_i(t)$ (where tensile strain is considered positive) and its gradient ${\partial \epsilon_i}/{\partial a_i}(t)$, which are influenced by the interaction dynamics (Eqs.~\ref{eq: function f} and \ref{eq: function g}), the reaction force $r_i(t)$ and its gradient ${\partial r_i}/{\partial a_i}(t)$ are incorporated as elements of the environmental state $z_{(i;0)}(t)$ surrounding the component object $i$ associated with the interaction object $j$. Additionally, to define the interaction conditions \textbf{\IntTwo{}} and \textbf{\IntThree{}}, the environmental state $z_{(i;0)}(t)$ includes the elastic strain energy and its gradient with respect to the cross-sectional area, denoted as
\begin{align}
  U^{z}_i(t)
 &= \dfrac{1}{2} 
    \dfrac{{e_ia_i(t)}/{l_i}}
          {\mleft(\sum_{k\in C}
                  \mleft({e_ka_k(t)}/{l_k}\mright)
          \mright)^2}\ p^2,
  \label{eq: cs Uz}
  \\
  \dfrac{\partial U^{z}_i}{\partial a_i}(t)
 &= \dfrac{1}{2} 
    \dfrac{\mleft({e_i}/{l_i}\mright) 
           \mleft({e_ja_j(t)}/{l_j} - {e_ia_i(t)}/{l_i} \mright)}
          {\mleft(\sum_{k\in C}
                  \mleft({e_ka_k(t)}/{l_k}\mright)\mright)^3}\ p^2
  \quad (j \neq i),
  \label{eq: cs Uzgrad}
\end{align}
expressed using the external force $p$. Thus, the environmental state $z_{(i;0)}(t)$ is defined as
\begin{align}
  z_{(i;0)}(t)
  = \mleft(r_i(t), 
           \dfrac{\partial r_i}{\partial a_i}(t), 
           \dfrac{\partial U^{z}_i}{\partial a_i}(t) 
    \mright)
    \in S^{z}_{(i;0)}=\mathbb{R}^3.
  \label{eq: cs beta_(i;0)}
\end{align}

The energy function that forms the energy landscape of each component object is defined as the elastic strain energy expressed by its total internal state $s_i(t) = (x_i(t),y_i(t))$. Therefore, the energy function $ U_i[y_i(t)] $ (Eq.~\ref{eq: energy function}) and the energy change rate function $ J_i[s_i(t)] $ (Eq.~\ref{eq: energy gradient function}) are defined based on the strain $\epsilon_i(t)$ and its gradient with respect to the cross-sectional area ${\partial \epsilon_i}/{\partial a_i}(t)$, denoted as
\begin{align}
  U_i[y_i(t)](x_i)
 &= {\dfrac{1}{2}}{e_i}{l_i}{\epsilon_i(t)^2}{a_{i}},
 \label{eq: cs energy function}
  \\ 
  J_i[s_i(t)](\underline{\dot x}_i)
 &= \dfrac{\partial U_i}{\partial a_i}[y_i(t)](x_i(t))\underline{\dot a}_i
  \notag\\ 
 &= \mleft({\dfrac{1}{2}}{e_i}{l_i}\epsilon_i(t)^2 
           + {e_i}{l_i}{\epsilon_i(t)}{\dfrac{\partial \epsilon_i}{\partial a_i}(t)}{a_{i}(t)} 
    \mright)\underline{\dot a}_i.
  \label{eq: cs energygrad function}
\end{align}
In addition, the energy dissipation rate function $\Psi_i[s_i(t)]$ (Eq.~\ref{eq: cost function}) is defined based on Eq.~\eqref{eq: cs x_i} as
\begin{align}
  \Psi_i[s_i(t)](\underline{\dot x}_i)
 &= {\dfrac{1}{2}}{\omega_i(t)}{\underline{\dot a}_i^2}.
  \label{eq: cs cost function}
\end{align}
By using these definitions, the rate $\underline{\dot a}_i(t)$ of adaptive cross-sectional area changes is defined based on Eqs.~\eqref{eq: Urate} and \eqref{eq: gradient flow} as
\begin{align}
  \underline{\dot a}_i(t)
 &= -\dfrac{1}{\omega_i(t)}
     \dfrac{\partial U_i}{\partial a_i}[y_i(t)](x_i(t)).
  \label{eq: cs gradient flow}
\end{align}
Thus, each component object has the capability of decreasing its elastic strain energy $U_i[y_i(t)](x_i)$ at the rate $U^\text{rate}_i[s_i(t)](\underline{\dot x}_i(t))$ through the cross-sectional area change rate $\underline{\dot x}_i(t) = \underline{\dot a}_i(t)$ following the minimization condition of Eq.~\eqref{eq: gradient flow}.

\subsection{Set up of interactions among living components}
Next, by defining the actual rate ${\dot x}_i(t)$ of the internal state $x_i(t)$ through interaction dynamics among the component objects, the rate $\dot y_i(t)$ that induces the temporal changes in the energy landscape formed by the function $U_i[y_i(t)]$, and the rate in $s_i(t)$ that induces the temporal changes in the energy rate landscape formed by the function $U^\text{rate}_i[s_i(t)]$, which are illustrated in \figurename~\ref{fig: view}(c), we set the functions $f_{(i;0)}$ and $g_{i} \ (i = 1,2)$ (Eqs.~\ref{eq: function f} and \ref{eq: function g}) to represent the dynamics of interactions among component objects for each interaction condition.

First, for simplicity, we employ the same function $f_{(i;0)}$ across all interaction conditions (\textbf{\IntOne{}}, \textbf{\IntTwo{}}, \textbf{\IntThree{}}). Thus, under all interaction conditions, the rate $\dot z_{(i;0)}(t)$ (Eq.~\ref{eq: function f}) of the environmental state of the component object $i$ includes the rates of the reaction force $r_i(t)$ (Eq.~\ref{eq: cs real Reaction force}), reaction force gradient $\mleft({\partial r_i}/{\partial a_i}\mright)(t)$ (Eq.~\ref{eq: cs real Reaction force gradient}), and the elastic strain energy gradient ${\partial U^{z}_i}/{\partial a_i}(t)$ (Eq.~\ref{eq: cs Uzgrad}), each in response to the cross-sectional area rate in adaptive behavior $(\underline{\dot a}_1(t),\underline{\dot a}_2(t))$, respectively denoted as
\begin{align}
  \dot r_i(t)
  &= -\dfrac{\sum_{k,m\in C,\ m \neq k} 
             \mleft({{r_m(t)}{e_k}{\underline{\dot a}_{k}(t)}}/{l_k}\mright)}
            {\sum_{k\in C}
             \mleft({e_k}{a_k(t)}/{l_k}\mright)}\ p,
  \\
  \dot{\mleft(\dfrac{\partial r_i}{\partial a_i}\mright)}(t)
  &= \dfrac{e_i}{l_i}
     \dfrac{\sum_{k\in C} 
            \mleft({r_k(t)}{{e_j}{\underline{\dot a}_{j}(t)}}/{l_j}\mright) 
            + 2{r}_{j}(t){e_i\underline{\dot a}_{i}(t)}/{l_i(i)}}
           {\mleft(\sum_{k\in C}\mleft({e_ka_k(t)}/{l_k}\mright)\mright)^2}
           \ p \quad (j \neq i),
   \\
  \dot{\mleft(\dfrac{\partial U^{z}_i}{\partial a_i}\mright)}(t)
  &= (-1)^i
    \dfrac{e_i}{l_i}
    \dfrac{\sum_{k,m\in C,\ m \neq k}
           \mleft({r_k(t)+2r_m(t)}
                  {{e_k}{\underline{\dot a}_{k}(t)}}/{l_k}\mright)}
          {\mleft(\sum_{k\in C}
                  \mleft({e_ka_k(t)}/{l_k}\mright)\mright)^3}\ p^2.
  \label{eq: cs Uzgraddot}
\end{align}

Then, we define the function $g_{i}$ for each interaction condition. For simplicity, we assume that the actual cross-sectional area rate $\dot a_i(t)$ through interaction dynamics is identical to the adaptive cross-sectional area rate $\underline{\dot a}_i(t)$ ($\dot a_i(t) = \underline{\dot a}_{i}(t)$, or equivalently, $\dot x_i(t) = \underline{\dot x}_{i}(t)$). Under this simplification, each interaction condition defines the temporal changes in the internal state $y_i(t) = ({\epsilon_i}(t),{{\partial \epsilon_i}/{\partial a_i}}(t),\omega_i(t))$ induced by the function $ g_{i}$.

\subsubsection{\textbf{\IntOne{}}: Global adaptive interaction}
To define the function $g_{i}\ (i=1,2)$ under the interaction condition \textbf{\IntOne{}}, we associate the adaptive behavior of the integrated living system with the function $g_{i}$ (\figurename~\ref{case study}(b)). For simplicity, we assume that the weight $\omega_i(t)$ of the energy dissipation rate in adaptive behavior remains constant, that is, $\dot \omega_i(t) = 0$. An energy function representing the elastic strain energy stored in the integrated system under the external force $p$, energy change rate function, energy dissipation rate function, and net energy change rate function are respectively defined as
\begin{align}
  U_0[y_1(t),y_2(t)](x_1,x_2)
 &= \dfrac{1}{2} 
    \dfrac{p^2}
          {\sum_{k\in C}
           \mleft({e_ka_k(t)}/{l_k}\mright)
          },
  \\
  J_0[s_1(t),s_2(t)](\underline{\dot x}_1,\underline{\dot x}_2)
 &= \displaystyle\sum_{k\in C} 
    {\dfrac{\partial U_0}{\partial a_k}[y_1(t),y_2(t)](x_1(t),x_2(t))\underline{\dot a}_k},
  \\
\Psi_0[s_1(t),s_2(t)](\underline{\dot x}_1,\underline{\dot x}_2)
 &= \sum_{k\in C}
    \Psi_k[s_k(t)](\underline{\dot x}_i),
  \\
  U^\text{rate}_0[s_1(t),s_2(t)](\underline{\dot x}_1,\underline{\dot x}_2)
 &= \mleft(J_0[s_1(t),s_2(t)] + \Psi_0[s_1(t),s_2(t)]\mright)(\underline{\dot x}_1,\underline{\dot x}_2).
  \label{eq: cs U0}
\end{align}
The cross-sectional area rates $(\underline{\dot a}_1(t),\underline{\dot a}_2(t))(=(\dot a_1(t),\dot a_2(t)))$ are assumed to satisfy the generalized gradient flow condition based on the function $U^\text{rate}_0[s_1(t),s_2(t)]$, denoted as
\begin{align} 
  (\underline{\dot a}_1(t),\underline{\dot a}_2(t))
  \ \text{minimizes}\ U^\text{rate}_0[s_1(t),s_2(t)].
  \label{eq: cs total_gradient_flow}
\end{align}
For the cross-sectional area rate $\underline{\dot a}_i(t)$ of each component object $i$ to simultaneously satisfy the generalized gradient flow condition based on the function $U^\text{rate}_i[s_i(t)]$, this condition requires that the reaction force $r_i(t)$ among its environmental states be reflected in its internal state whereas the reaction force gradient ${\partial r_i}/{\partial a_i}(t)$ not be.
Consequently, the strain $\epsilon_i(t)$ and its gradient with respect to the cross-sectional area ${\partial \epsilon_i}/{\partial a_i}(t)$ are given by
\begin{align}
  \epsilon_i(t) &= -\dfrac{r_i(t)}{e_ia_i(t)},
  \label{eq: cs epsilon IntOne}
  \\ 
  \dfrac{\partial \epsilon_i}{\partial a_i}(t) &= \dfrac{r_i(t)}{e_ia_i(t)^2}.
  \label{eq: cs epsilongrad IntOne}
\end{align}
Thus, under the interaction condition \textbf{\IntOne{}}, which satisfies these equations, the strain rate and strain gradient rate of component object $i$, defined by the function $g_{i}$ are derived as
\begin{align}
  \dot \epsilon_i(t)
 &= -\dfrac{{\dot r_i(t)}{a_i(t)} - {r_i(t)}{\underline{\dot a}_i(t)}}
           {{e_i}{a_i(t)^2}},
 \label{eq: cs epsilondot IntOne}
 \\
  \dot{\mleft(\dfrac{\partial \epsilon_i}{\partial a_i}\mright)}(t)
 &= \dfrac{{\dot r_i(t)}{a_i(t)} - 2{r_i(t)}{\underline{\dot a}_i(t)}}{{e_i}{a_i(t)^3}}.
 \label{eq: cs epsilongraddot IntOne}
\end{align}

\subsubsection{\textbf{\IntTwo{}}: Energy landscape modified interaction}
The interaction condition \textbf{\IntTwo{}} defines function $g_{i}$ so that dynamics of the energy landscape, instead of the adaptive behavior of the integrated living system, specifies the emergence of a function of the integrated living system (\figurename~\ref{case study}(c)). In \textbf{\IntTwo{}}, not only the reaction force $r_i(t)$, but also its gradient ${\partial r_i}/{\partial a_i}(t)$ is reflected in the internal state of each component object. Accordingly, the strain gradient ${\partial \epsilon_i}/{\partial a_i}(t)$ and its rate $\dot{\mleft({\partial \epsilon_i}/{\partial a_i}\mright)}(t)$ are respectively modified from Eqs.~\eqref{eq: cs epsilongrad IntOne} and \eqref{eq: cs epsilongraddot IntOne} to
\begin{align}
  \dfrac{\partial \epsilon_i}{\partial a_i}(t)
 &= \dfrac{r_i(t)}{e_ia_i(t)^2} - \dfrac{{\partial r_i}/{\partial a_i}(t)}{e_ia_i(t)},
  \label{eq: cs epsilongrad IntTwo}
  \\
  \dot{\mleft(\dfrac{\partial \epsilon_i}{\partial a_i}\mright)}(t)
 &= \dfrac{\dot {r_i}(t) a_i(t) - 2r_i(t) \underline{\dot a}_i(t)}{e_ia_i(t)^3} 
 \notag\\
    &- \dfrac{\dot{\mleft({\partial r_i}/{\partial a_i}\mright)}(t) a_i(t) 
             - {\mleft({\partial r_i}/{\partial a_i}\mright)(t)}
               {\underline{\dot a}_i(t)}}
            {e_ia_i(t)^2}
  \label{eq: cs epsilongraddot IntTwo}
\end{align}
(\figurename~\ref{case study}(c)). As a result, the energy gradient ${\partial U_i}/{\partial a_i}[y_i(t)](x_i(t))$ and its rate $\dot{({\partial U_i}/{\partial a_i})}(t) $ of each component object coincide with ${\partial U^{z}_i}/{\partial a_i}(t)$ (Eq.~\ref{eq: cs Uzgrad}) and its rate $\dot{({\partial U^{z}_i}/{\partial a_i})}(t)$ (Eq.~\ref{eq: cs Uzgraddot}), respectively.
Therefore, the signs of the cross-sectional area rate $\underline{\dot a}_i(t)$ (Eq.~\ref{eq: cs gradient flow}) and the energy gradient rate $\dot{({\partial U_i}/{\partial a_i})}(t)$ hold antisymmetry between the component objects, respectively expressed as
\begin{align}
  \sum_{k\in C}
    \mleft(\dfrac{\omega_k(0)}{{e_k}/{l_k}}\underline{\dot a}_k(t)\mright)
  &= \sum_{k\in C}
    \mleft(-\dfrac{{l_k}}{{e_k}}\dfrac{\partial U_k}{\partial a_k}[y_k(t)](x_k(t))\mright)
  =0,
  \label{eq: cs anti1 IntTwo}
  \\
  \sum_{k\in C}
  \mleft(\dfrac{\omega_k(0)}{{e_k}/{l_k}}\underline{\ddot a}_k(t)\mright)
  &= \sum_{k\in C}
    \mleft(-\dfrac{{l_k}}{{e_k}}\dot{\mleft(\dfrac{\partial U_k}{\partial a_k}\mright)}(t)\mright)
  =0.
  \label{eq: cs anti2 IntTwo}
\end{align}
Thus, under the interaction condition \textbf{\IntTwo{}}, the integrated living system exhibits a function that makes the symmetry between the signs of the cross-sectional area rate $ {\dot a}_i(t) (=\underline{\dot a}_i(t)) $ and its acceleration $ {\ddot a}_i(t) (=\underline{\ddot a}_i(t)) $ consistent across the component objects.

\subsubsection{\textbf{\IntThree{}}: Energy rate landscape modified interaction}
\begin{figure}[h]
  \centering
  \includegraphics[width=\columnwidth]{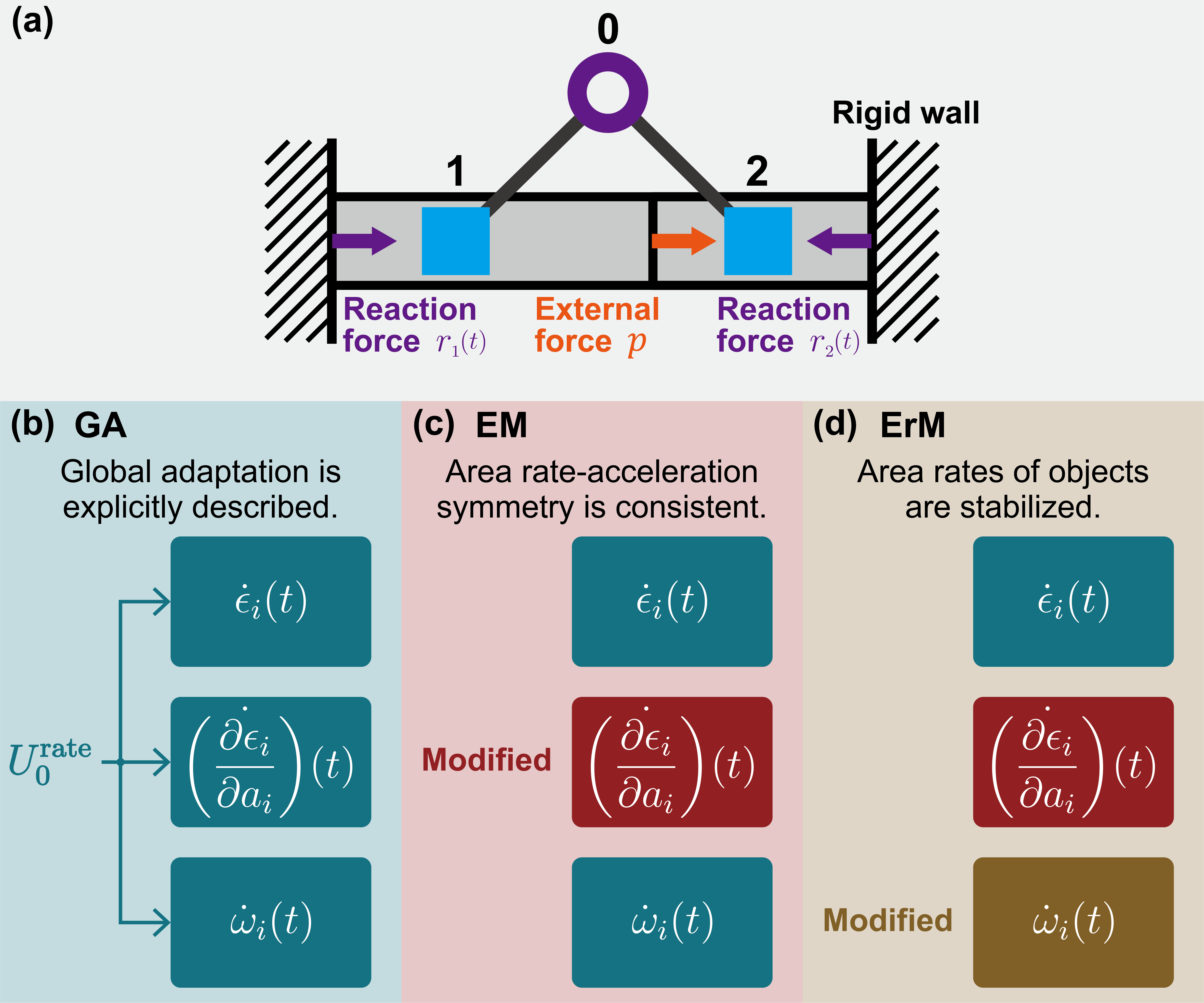}
  \caption{Set up of the case studies.
    (a) System considered in the case studies: the component objects 1 and 2 correspond to living components that behave as elastic bars and mechanically interact via the interaction object 0 under the external force $p$ at their joint; each component object experiences the reaction force $r_i(t)$ at the fixed point on the rigid wall.
    (b) Interaction condition \textbf{\IntOne{}}: Global adaptive interaction.
    (c) Interaction condition \textbf{\IntTwo{}}: Energy landscape modified interaction.
    (d) Interaction condition \textbf{\IntThree{}}: Energy rate landscape modified interaction.
    }
  \label{case study}
\end{figure}

The interaction condition \textbf{\IntThree{}} defines function $g_{i}$ so that dynamics of the energy rate landscape, instead of the adaptive behavior of the integrated living system, specifies the emergence of a function of the integrated living system (\figurename~\ref{case study}(d)). In \textbf{\IntThree{}}, in addition to the strain gradient rate $\dot{\mleft({\partial \epsilon_i}/{\partial a_i}\mright)}(t)$, the rate of the weight $\omega_i(t)$ of the energy dissipation rate is modified from $\dot \omega_i(t) = 0$ to be dependent on the elastic strain energy gradient rate $\dot{\mleft({\partial U^{z}_i}/{\partial a_i}\mright)}(t)$, expressed as
\begin{align}
  \dot \omega_i(t)
  = \omega_i(t)
    \dfrac{\dot{\mleft({\partial U^{z}_i}/{\partial a_i}\mright)}(t)}
          {{\partial U^{z}_i}/{\partial a_i}(t)}.
  \label{eq: cs omegadot}
\end{align}
(\figurename~\ref{case study}(d)).
Consequently, the cross-sectional area acceleration $ \underline{\ddot a}_i(t) $ (Eq.~\ref{eq: cs gradient flow}) for each component object satisfies
\begin{align}
  \underline{\ddot a}_i(t)
  = \dfrac{1}{\omega_i(t)^2}
  \mleft(\dot\omega_i(t)\dfrac{\partial U^{z}_i}{\partial a_i}(t)
  - \omega_i(t)\dot{\mleft(\dfrac{\partial U^{z}_i}{\partial a_i}(t)\mright)}\mright) = 0.
  \label{eq: cs zero IntThree}
\end{align}
Thus, under the interaction condition \textbf{\IntThree{}}, the integrated living system exhibits a function that stabilizes the cross-sectional area rate $\underline{\dot a}_i(t)$ of each component object.

\subsection{Results}
\begin{figure}[h]
  \centering
  \includegraphics[width=\columnwidth]{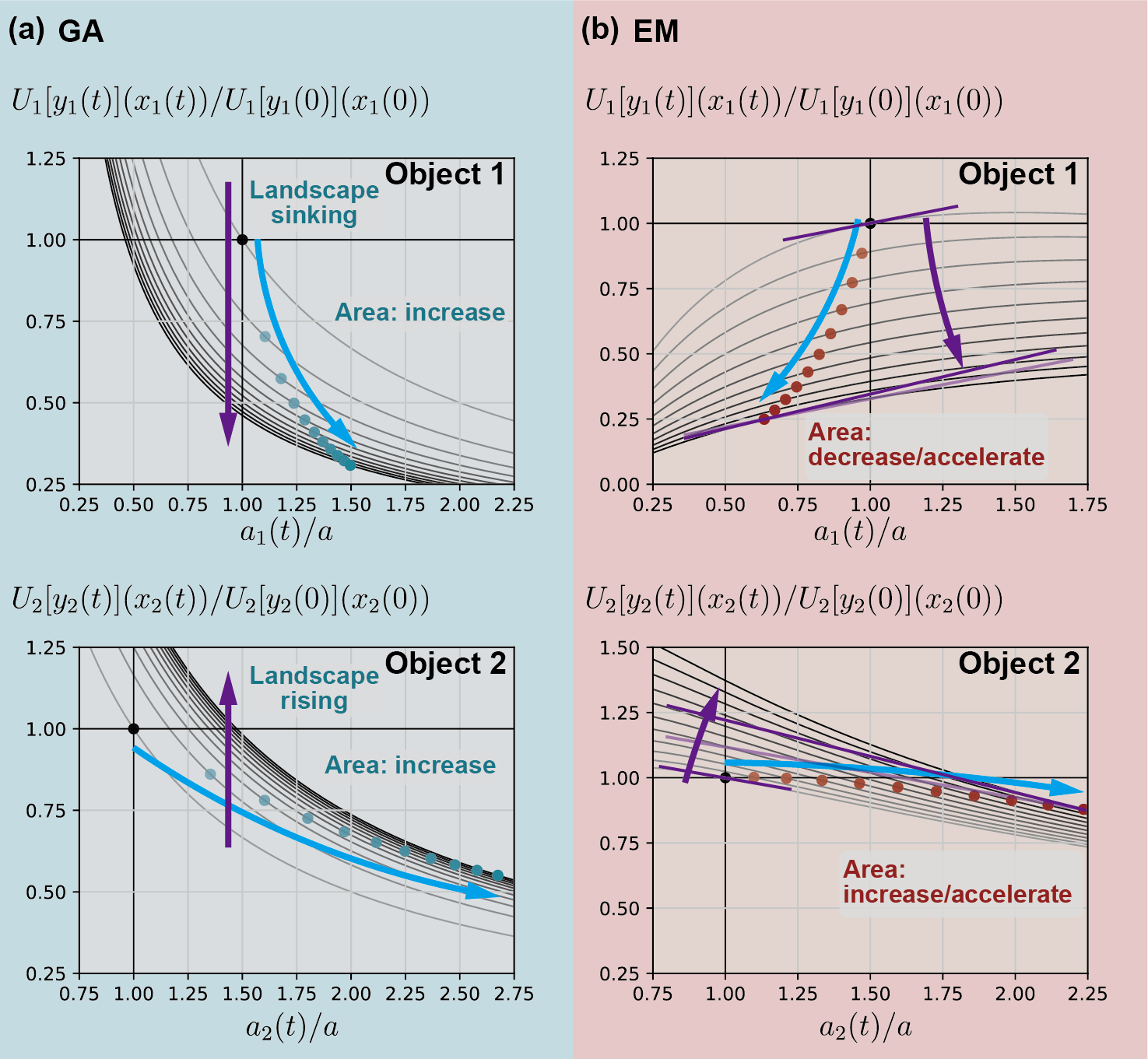}
  \caption{Changes in the sectional areas, energies, and energy landscapes over time 
  (a) under \textbf{\IntOne{}}: Global adaptive interaction, and
  (b) under \textbf{\IntTwo{}}: Energy landscape modified interaction.}
  \label{case study1}
\end{figure}

\begin{figure}[h]
  \centering
  \includegraphics[width=\columnwidth]{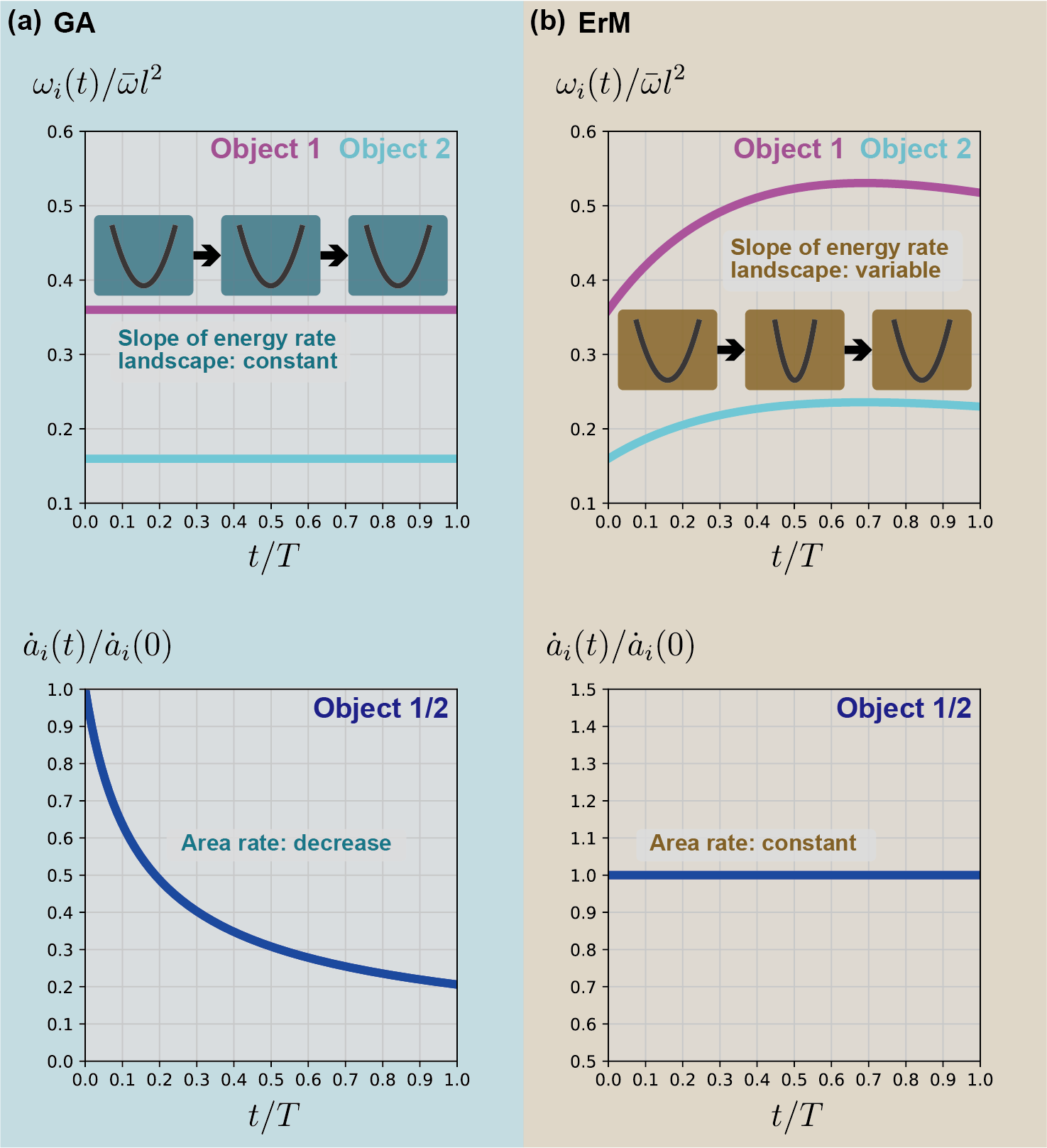}
  \caption{Changes in the slopes of the energy rate landscapes and sectional area rates over time 
  (a) under \textbf{\IntOne{}}: Global adaptive interaction, and 
  (b) under \textbf{\IntThree{}}: Energy rate landscape modified interaction.}
  \label{case study2}
\end{figure}

Under each interaction condition (\textbf{\IntOne{}}, \textbf{\IntTwo{}}, \textbf{\IntThree{}}) defined in the previous section, we analyzed the behavior of the cross-sectional areas, as well as the energy or energy rate landscape of each component over time (\figurename~\ref{case study1}, \ref{case study2}). Here, the time interval for analyzing the system dynamics was defined as $t \in [0, T]$ by using a constant $T$. For simplicity, the initial values of the cross-sectional area $a_i(0)$ and Young modulus $e_i$ were consistent across the components, given by $a_i(0) = a \ (=\text{const.})$ and $e_i = e \ (=\text{const.})$, respectively. To make the initial cross-sectional area rates $\dot a_i(0)$ nonzero under \textbf{\IntTwo{}}, the natural lengths $l_i$ are set as $l_1 = 0.6\ l,\ l_2 = 0.4\ l$ by using the constant $l \ (=l_1+l_2)$, which represents the distance between the rigid walls. Additionally, the initial weight $\omega_i(0)$ of the energy dissipation rate in adaptive behavior was consistent across the components, given by $\omega_i(0) = \bar{\omega} l_i^2$ by using a constant $\bar{\omega}$ and the length $l_i$.

In \figurename~\ref{case study1}, to confirm the emergence of the function of the integrated living system under the interaction condition \textbf{\IntTwo{}} through the energy landscape dynamics of the component objects, we illustrate the dynamics of the cross-sectional area $a_i(t)$, the energy $U_i[y_i(t)](x_i(t))$, and the energy landscape formed by the energy function $U_i[y_i(t)]$ under the interaction conditions \textbf{\IntOne{}} and \textbf{\IntTwo{}} in \figurename~\ref{case study1}(a) and (b), respectively. Here, the cross-sectional area $a_i(t)$ and the energy $U_i[y_i(t)](x_i(t))$ were normalized by using their initial values, $a_i(0) = a$ and $U_i[y_i(0)](x_i(0))$.

Under the interaction condition \textbf{\IntOne{}}, because the energy gradient ${\partial U_i}/{\partial a_i}[y_i(t)](x_i(t))$ (Eq.~\ref{eq: cs energygrad function}) under substitution of the strain $\epsilon_i(t)$ (Eq.~\ref{eq: cs epsilon IntOne}) and its gradient ${\partial \epsilon_i}/{\partial a_i}(t)$ (Eq.~\ref{eq: cs epsilongrad IntOne}) is always positive, as illustrated by the trajectories of the points (blue arrow) in \figurename~\ref{case study1}(a), the cross-sectional areas $a_i(t)$ of both component objects increased over time. For component object 1, in response to the interaction dynamics, the energy landscape formed by the energy function $U_1[y_1(t)]$ (Eq.~\ref{eq: cs energy function}) sank over time, as illustrated by the trajectory of the curve (purple arrow) in \figurename~\ref{case study1}(a). Consequently, due to the effects of cross-sectional area increase and landscape sinking, the energy $U_1[y_1(t)](x_1(t))$ decreased. On the other hand, for component object 2, although the energy landscape rose, the effect of the increasing cross-sectional area outweighed the effect of the landscape rising. As a result, similar to component object 1, the energy $U_2[y_2(t)](x_2(t))$ decreased.

Under the interaction condition \textbf{\IntTwo{}}, as illustrated by the trajectory of the points (blue line) in \figurename~\ref{case study1}(b), the cross-sectional area rate $\dot a_1(t)$ of component object 1 turned into negative. Furthermore, as illustrated by the trajectory of the landscape (purple curve) in \figurename~\ref{case study1}(b), the energy gradient ${\partial U_1}/{\partial a_1}[y_1(t)](x_1(t))$ decreased due to the landscape changes, leading to a reduction in the cross-sectional area acceleration ${\ddot a}_1(t)$ of component object 1. On the other hand, for component object 2, due to the antisymmetric relationship of the signs of the cross-sectional area rates $\dot a_i(t) \ (i = 1, 2)$ between objects (Eq.~\ref{eq: cs anti1 IntTwo}), the cross-sectional area rate $\dot a_2(t)$ increased. Furthermore, because the sign of the energy gradient rate $\dot{({\partial U_i}/{\partial a_i})}(t)$ is also antisymmetric between objects (Eq.~\ref{eq: cs anti2 IntTwo}), the energy gradient ${\partial U_2}/{\partial a_2}[y_2(t)](x_2(t))$ increased with the landscape change, leading to an increase in the cross-sectional area acceleration ${\ddot a}_2(t)$ of component object 2. Thus, under the interaction condition \textbf{\IntTwo{}}, the function that makes the symmetry between the sign of the cross-sectional area rate ${\dot a}_i(t)$ and its acceleration ${\ddot a}_i(t)$ consistent across the component objects was confirmed through the energy landscape dynamics of the component objects.

In \figurename~\ref{case study2}, to confirm the emergence of the function of the integrated living system under the interaction condition \textbf{\IntThree{}} through the energy rate landscape dynamics of the component objects, we illustrated the dynamics of the weight $\omega_i(t)$ of the energy dissipation rate in adaptive behavior and the cross-sectional area rate $\dot a_i(t)$ under the interaction conditions \textbf{\IntOne{}} and \textbf{\IntThree{}} in \figurename~\ref{case study2}(a) and (b), respectively. Here, the time $t$ was normalized by using the constant $T$. Additionally, the weight $\omega_i(t)$ of the energy dissipation rate and the cross-sectional area rate $ \dot a_i(t) $ are normalized by using their respective initial values $\omega_i(0) = \bar{\omega} l^2_i$ and $\dot a_i(0)$, respectively.

Under the interaction condition \textbf{\IntOne{}}, as shown in \figurename~\ref{case study2}(a), the weight $\omega_i(t)$, that is, the slope of the energy rate landscape remained constant since the rate $\dot \omega_i(t)$ of the weight was set to be constant. Additionally, since the energy gradient $-{\partial U_i}/{\partial a_i}[y_i(t)](x_i(t))$ decreased as the cross-sectional area increased, the cross-sectional area rate $ \dot a_i(t)=-\omega_i(t)^{-1}{\partial U_i}/{\partial a_i}[y_i(t)](x_i(t)) $ decreased over time.

Under the interaction condition \textbf{\IntThree{}}, as the rate of the weight $\omega_i(t)$ was variable according to Eq.~\eqref{eq: cs omegadot}, the slope of the energy rate landscape changed over time, as shown in \figurename~\ref{case study2}(b). As a result, as derived from Eq.~\eqref{eq: cs zero IntThree}, the cross-sectional area rate $\dot a_i(t)$ of each component object remained constant. Thus, under the interaction condition \textbf{\IntThree{}}, the function that stabilizes the cross-sectional area rate $\dot a_i(t)$ of each component object was confirmed through the energy rate landscape dynamics of each component object.

Through the above case studies under three interaction conditions, the influence of interaction dynamics on the emergence of functions of an integrated living system was demonstrated to be analyzed by specifying the function of the integrated living system based on the dynamics of the energy landscapes and energy rate landscapes of individual component objects, not only under interaction conditions where the adaptive behavior of the integrated living system is explicitly defined (such as interaction condition \textbf{\IntOne{}} in the case studies) but also under interaction conditions where it is not (such as interaction conditions \textbf{\IntTwo{}} and \textbf{\IntThree{}} in the case studies).

\clearpage
\section{Discussion}
In this study, focusing on a living system composed of interacting living components, we developed a theoretical framework that embodies interaction dynamics, which is driven by the adaptive behavior of individual living components, based on temporal changes in energy landscapes. For the adaptive behavior of individual living components constituting a living system, we provided a formulation according to the generalized gradient flow of an energy landscape. Moreover, to express the dynamics of interactions among component objects due to their adaptive behavior, we formulated temporal changes in the environmental state $z_{(k;j)}(t)\ (j \in \mathcal{V}^{I}_{i},\ k \in \mathcal{V}^{C}_{j})$ surrounding each component object and in the total internal state $s_i(t) = (x_i(t), y_i(t))$ of each component object, each in response to the rate of adaptive internal state changes $\underline{\dot x}_{i}(t)$. Through this sequential formulation, the dynamics of interactions among component objects was represented based on temporal changes in the energy and energy rate landscapes respectively formed by the functions $U_i[y_i(t)]$ and $U^\text{rate}_i[s_i(t)]$.

Through the case studies using the model of mechanically interacting tissues, we demonstrated that, under an interaction condition that explicitly sets the adaptive behavior of the integrated living system according to the function $U_0[y_1(t),y_2(t)](x_1,x_2)$ (\textbf{\IntOne{}}), the energy landscape-based representation of interaction dynamics allowed us to associate the dynamics with the emergence of functions of the integrated living system. Furthermore, even under the interaction conditions without explicitly setting the adaptive behavior of the integrated living system (\textbf{\IntTwo{}}, \textbf{\IntThree{}}), setting the temporal changes in the energy and energy rate landscapes could represent the emergence of functions of the integrated living system resulting from the interaction dynamics. Therefore, our theoretical framework demonstrated that temporal changes in applied forces due to morphological changes of tissues determine system functions even without considering system-level adaptive behavior. These findings highlight that expressing interaction dynamics based on temporal changes in energy and energy rate landscapes offers a powerful theoretical framework for understanding how system functions emerge from adaptive behavior at the component level.

In this study, we conceptualized temporal changes in the surrounding environment of each living component as dynamics of interactions among components driven by their adaptive behavior and formulated the interaction dynamics based on temporal changes in energy and energy rate landscapes. Applying this landscape-based theoretical approach to actual living phenomena would contribute to understanding the mechanisms underlying the emergence of functions of an integrated living system where the environment of each component temporally changes along with the dynamical interactions among components. Examples of temporally changing environments under component interactions include the microenvironment reconstructed during epithelial-mesenchymal transition \citep{youssef_epithelialmesenchymal_2024}, the tumor microenvironment formed through signal exchange between cancer and non-cancerous cells \citep{de_visser_evolving_2023}, and the inflammatory or regenerative microenvironment shaped by the molecular secretion of senescent cells (senescence-associated secretory phenotype) \citep{paramos-de-carvalho_right_2021, saito_role_2024}. By applying our theoretical framework to these environments, diverse living phenomena would be understood from a unified perspective based on the adaptive behavior of individual living components and the resulting interaction dynamics.

Furthermore, not only living systems but also artificial systems such as swarm robots \citep{yang_survey_2022} and soft robots \citep{baines_robots_2024, van_laake_bio-inspired_2024}, which exhibit behavior inspired by living systems, can benefit from our theoretical framework for modeling interactions among components. When designing these artificial systems, considering only the behavior of individual components may require a detailed design of the behavior of each component over time to align with an intended system-wide behavior. By applying our theoretical framework to artificial system design, instead of increasing the complexity of designing individual component behavior, it can become possible to design dynamics of interactions among components in response to their adaptive behavior. Consequently, this approach is expected to enable the emergence of integrated system functions while maintaining simple design requirements for individual components.

In this study, we assumed that each component object corresponds to a single living component, and thus, the interaction objects did not support interactions at the level of clusters composed of multiple living components. To overcome this limitation, incorporating hypergraph theory could be an effective approach. In a hypergraph, hyperedges are defined as subsets of a set of nodes, and thus, each hyperedge can be interpreted as a cluster of nodes. Therefore, introducing a hypergraph where each node corresponds to a living component and assigning hyperedges to component objects would allow a single component object to treat clusters of living components. This approach enables, for example, both the energy landscapes representing the adaptive behavior of an individual living component and of a cluster containing that component to change in response to interaction dynamics and, consequently, this extension allows for more flexible and appropriate modeling of living system dynamics.

Furthermore, mathematical refinements represented by the above extensions would facilitate the integration of this theoretical framework with other theories that focus on interactions among living components and enable a broader mathematical perspective for understanding how interaction dynamics influences the emergence of functions of an integrated living system. To describe living system dynamics under component interactions, various theories have been proposed based on, for example, the Hopfield network \citep{herron_robust_2023}, the information thermodynamics \citep{borsley_membrane_2024}, and the game theory \citep{fujimoto_game-theoretical_2024}. Applying our theoretical framework to these existing theories, which are available for modeling living systems, could enable these theories to explicitly describe the effects of interaction dynamics, which have thus far only been considered implicitly, by using temporal changes in energy and energy rate landscapes.

\section*{Acknowledgment}
This work was supported by Grant-in-Aid for Scientific Research (A) (JP25H01219) and (A) (JP20H00659) from Japan Society for the Promotion of Science (JSPS); JST-CREST (JPMJCR22L5) and JST SPRING (JPMJSP2110) from Japan Science and Technology Agency (JST).

\section*{Data availability}
The data that support the findings of this article are openly available at KURENAI (\url{https://doi.org/10.57723/kds590896}).

\section*{CRediT authorship contribution statement}
\textbf{Ryunosuke Suzuki:} Writing - review and editing, Writing - original draft, Visualization, Software, Resources, Methodology, Investigation, Funding acquisition, Formal analysis, Conceptualization. \textbf{Taiji Adachi:} Writing - review and editing, Supervision, Resources, Project administration, Funding acquisition, Conceptualization.

\bibliographystyle{elsarticle-harv.bst}
\bibliography{IntJ.bib}

\end{document}